          [1999/12/02 1.01 (PWD)]
\documentclass[a4paper,twocolumn]{esapub} 

\usepackage{epsfig} 		
\usepackage{natbib}

\title{AGN: The High-Energy Status before INTEGRAL}
\author{W.~Collmar}
\affil{Max-Planck-Institut f\"ur extraterrestrische Physik, Garching, Germany}

\def \deg     {$^o$}

\def \gray    {$\gamma$-ray}
\def \grays   {$\gamma$-rays}

\def \msun{{\rm M}_\odot}

\begin{document}


\maketitle

\begin{abstract}
Much progress in the knowledge of the high-energy emission
($>$50~keV) from Active Galactic Nuclei has been made during the 
last decade, predominately by the experiments aboard 
the Compton Gamma-Ray Observatory. After the end of 
the CGRO mission and prior to the upcoming INTEGRAL mission 
the current status of high-energy emission properties and scenarios
is summarized for different AGN classes. In addition, 
prospects for INTEGRAL observations of AGN are given.     

\end{abstract}

\section{Introduction}

Considerable progress on our knowledge about 
Active Galactic Nuclei (AGN) was made during the last
decade, e.g., in the optical by the Hubble Space Telescope, by several 
new missions in the X-ray domain (e.g., ROSAT, ASCA). 
A lot of progress was also achieved at even higher photon energies 
in the \gray\ domain, in particular by the experiments aboard the 
Compton Gamma-Ray Observatory (CGRO)
($\sim$50~keV to $\sim$20~GeV) and by improved ground-based Cherenkov 
telescopes at energies above several hundreds of GeV. 
In this paper I shall try to briefly summarize the current status of 
our knowledge on AGN at \gray\ energies by describing their 
observational properties and the anticipated emission scenarios. 
Because the launch of the INTEGRAL mission is on the horizon, 
I shall also mention some prospects for the INTEGRAL experiments. 

Roughly 3\% of all
galaxies are classified as being \lq active\rq. 
Although there is no overall agreement amongst astronomers about
a precise definition of active galaxies, one common criterion
is the generation of 
large luminosities in small core regions. In many cases
(e.g., quasars) the active and bright core overshines the remaining
galaxy by far. 
A large number of classes and sub-classes exist, which developed
historically and sometimes overlap in their parameters.

\begin{figure}[th]
\centering
\epsfig{figure=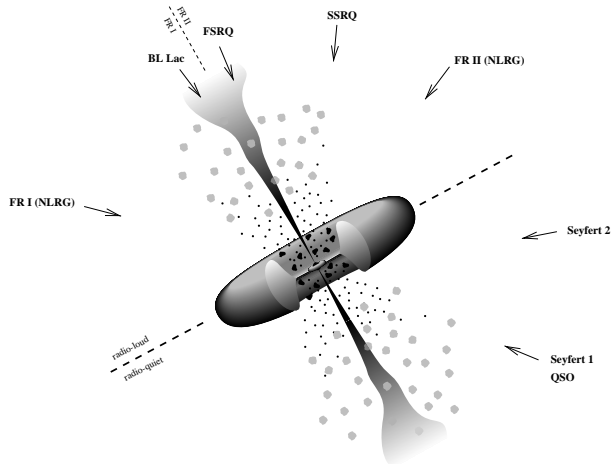,width=8cm,clip=}
\caption{
The \lq standard model\rq\ for AGN. The figure is adapted from
Urry \& Padovani (1995).
\label{fig:agn_scen}}
\end{figure}

The \lq unification-by-orientation\rq\ scenario assumes
that all AGN are intrinsically similar. 
At the very center there is a supermassive black hole 
($\sim$10$^6$ to $\sim$10$^{10}\,\msun$), which accretes matter from
an accretion disk. 
This central region is surrounded by an
extended molecular torus and moving gas clouds.  
In addition to these clouds a hot electron corona
populates the inner region.
In a radio-loud AGN a strong jet of relativistic particles
emanates perpendicular to the plane of the accretion disc.
The generation of such jets is still not 
understood; however, it is believed that strong magnetic
fields play a fundamental role.
The \lq unification-by-orientation\rq\ scenario assumes such a general
structure for all AGN. Depending on the spatial orientation with
respect to our line of sight we observe the different types of AGN
(Fig.~\ref{fig:agn_scen}; for a full discussion see e.g.,
Urry \& Padovani 1995).
If we look towards the \lq central engine\rq\ along
 the jet axis ($<$10\deg),
i.e., basically directly into the jet, we observe a blazar.
A regular quasar or a Seyfert-1 galaxy is observed if we look at 
an offset angle of the order of 30\deg, at which both the narrow-line
and broad-line regions are visible. At larger angular offsets the
broad-line region will be hidden by this extended molecular torus
giving rise to Seyfert-2 galaxies. A typical radio galaxy, 
showing two strong opposite jets, is observed at angles approximately
perpendicular to the jet axis. 

Prior to the launch of CGRO in April 1991
the extragalactic \gray\ sky was largely unknown.
Only the quasar 
3C~273 had been detected significantly as an emitter 
of $\gamma$-radiation at energies above 50~MeV by the 
COS-B satellite\index{COS-B} \citep{Swanenburg78}.
For three more sources the detection of \grays\ up
to $\sim$20~MeV had been reported.
However, these measurements resulted from short-duration
balloon flights and contained large systematic uncertainties.  
Now, after CGRO, about 100~AGN (the number depends on the assumed 
detection threshold level) are detected by the different CGRO instruments
at energies above 50~keV: $\sim$90 blazars, $\sim$25 Seyfert galaxies, 
and 1 radio galaxy. These detections opened the field of extragalactic
\gray\ astronomy and astrophysics and, in addition, provided an important 
step in the understanding of AGN physics in general.  
In the following I shall briefly summarize the \gray\ status for the 
different AGN types.  
  
\section{Blazars}
\subsection{\lq EGRET-type\rq\ Blazars}

The detection of many blazars at \gray\ energies above 
100~MeV by the EGRET experiment aboard CGRO finally opened 
the field of extragalactic \gray\ astronomy. 
Blazars are radio-loud quasars or BL~Lac objects which show
a flat radio spectrum, strong and rapid variability in optical and 
radio bands, and strong optical polarization. 
Although the blazar 3C~273 had been discovered at
these energies by COS-B 
about 25 years ago \citep{Swanenburg78}, the detection of so many
sources came as a surprise.
Roughly 90 blazars are now detected by EGRET 
\citep{Hartman99}, 10 of them are also observed at MeV \grays\ 
(0.75 - 30~MeV) by the COMPTEL experiment \citep{Collmar99a}, 
and also $\sim$10 by the OSSE experiment \citep{Kurfess96}, 
mainly at hard X-ray to soft \gray\ energies (50~keV to 500~keV). 
The term \lq EGRET-type\rq\ refers to 
blazars which are prominent in the energy range from $\sim$1~MeV 
to a few GeV. 
   
\begin{figure}[th]
\centering
\epsfig{figure=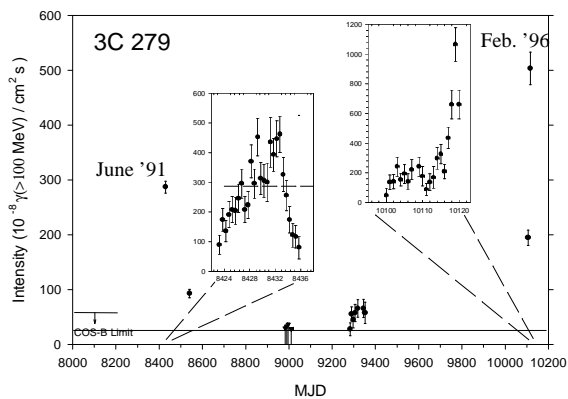,width=8cm,clip=}
\caption{
Long-term \gray\ light curve of the blazar 3C~279 at energies 
above 100~MeV \protect{\citep{Hartman97}}. Large flux 
variations are visible. The two flaring events in June 1991
and February 1996 are shown in the insets with improved time resolution 
of 0.5 days and 1 day, respectively.
Variability on time scales of 1 day is obvious.  
\label{fig:3c279_lc}}
\end{figure}

\begin{figure}[th]
\centering
\epsfig{figure=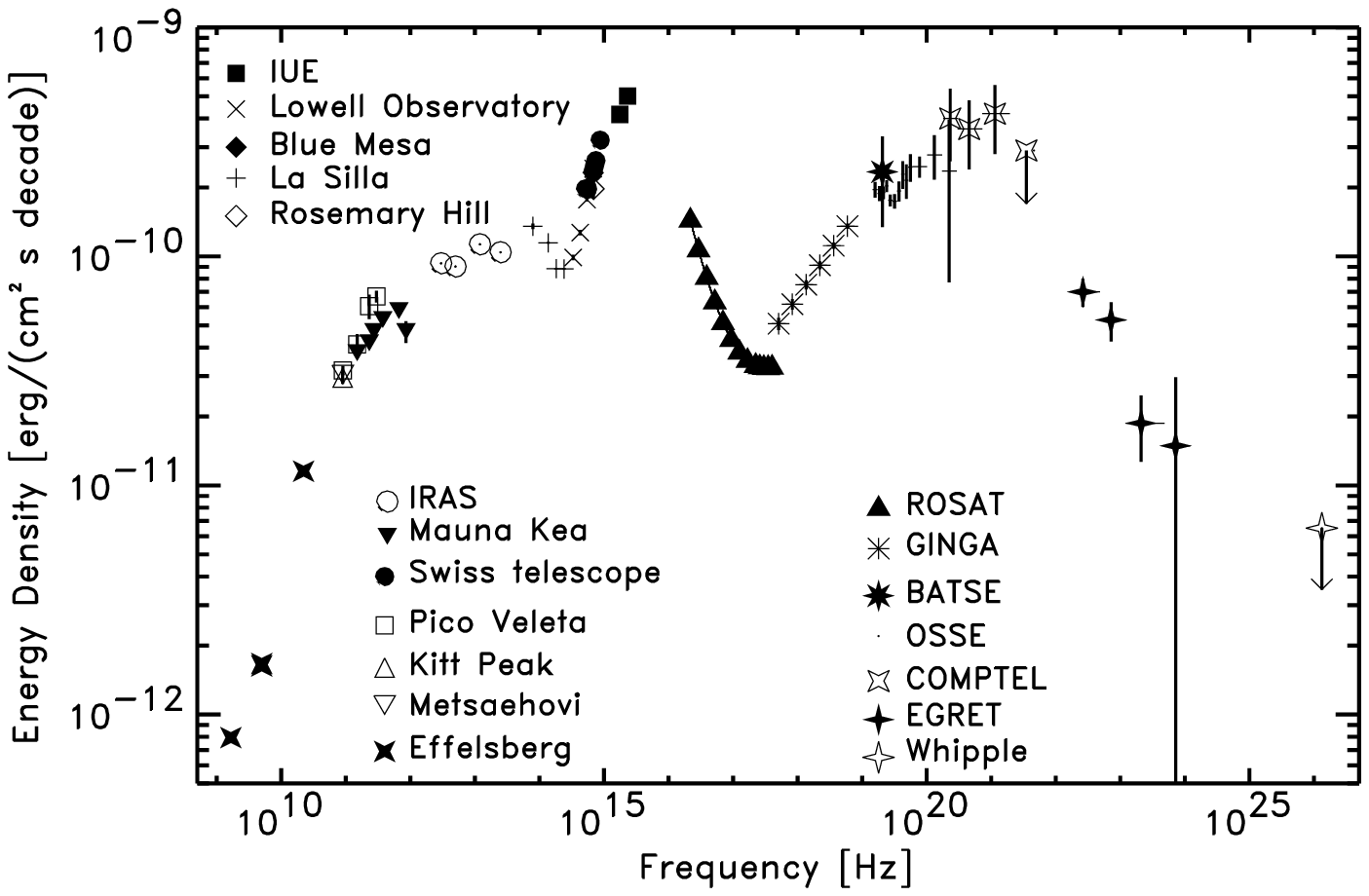,width=8cm,clip=}
\vspace*{-1.0cm}
\epsfig{figure=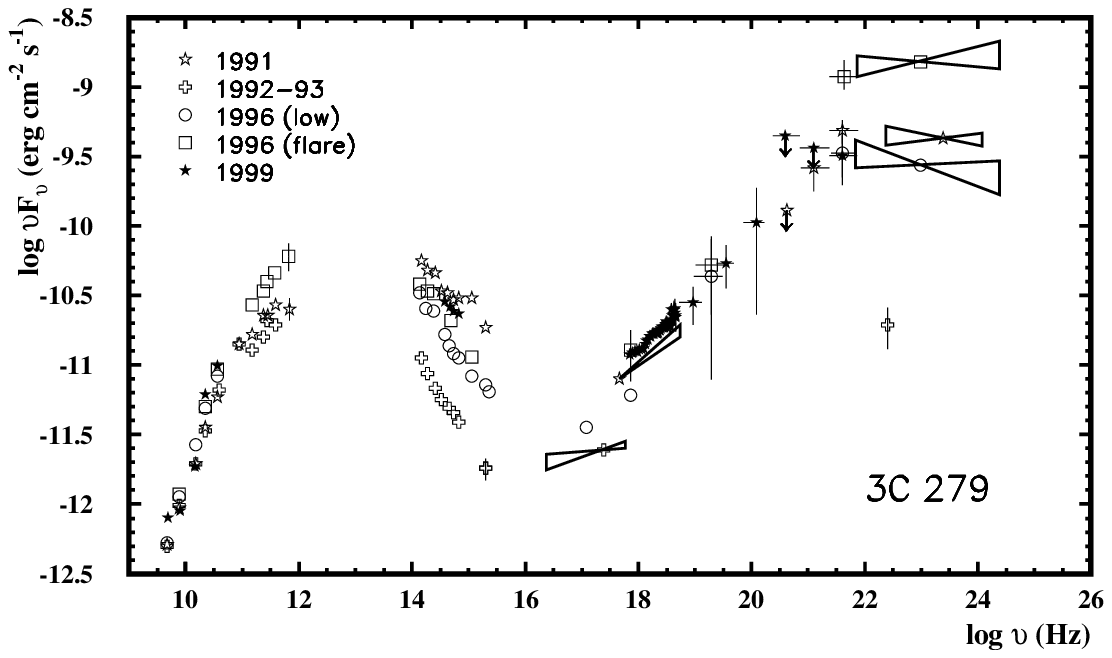,width=8cm,clip=}
\vspace*{0.3cm}
\caption{
Upper panel: Quasi-simultaneous broad-band spectrum of 3C~273 
\protect\citep{Lichti95}.
Several maxima are visible, which point to different emission 
components and processes (see text). 
\newline
Lower panel: 
Broad-band spectra of 3C~279 for different times and different 
\gray\ flux states \protect{\citep[from][]{Collmar00}}.
Only two emission maxima are visible which 
are interpreted to be due to two non-thermal emission processes: 
synchrotron radiation from radio to UV/X-rays and IC emission 
from UV/X-rays to \grays. The thermal emission components are probably
outshone by the dominance of the non-thermal ones. 
 \label{fig:blazars_spec}}
\end{figure}

The \gray\ fluxes of blazars are found to be highly time variable,
even on time scales as short as one day and less,  
showing flux variations up to a factor of $\sim$100.
Fig.~\ref{fig:3c279_lc}, as an example, shows the long-term light
curve of 3C~279 as 
observed by EGRET at energies above 100~MeV. 
Large flux variations as well as short-term variability is clearly 
seen. The individual CGRO instruments measure spectra 
which typically can be represented by power-law shapes. 
However, if the spectra are combined a spectral turnover from a 
softer shape at high-energy \grays\ ($>$100~MeV) to a harder shape 
at hard X-rays ($>$50~keV) becomes often visible. This spectral 
turnover occurs at MeV energies (as an example see
 Fig.~\ref{fig:blazars_spec}). 
By assuming isotropic emission, typical \gray\ luminosities of 
the order of 10$^{48}$~erg/s with maxima up to several 
10$^{49}$~erg/s are derived \citep[e.g.,][]{Mattox97,Collmar97}.
Keeping in mind that the bolometric luminosity of our \lq Milky Way\rq\ 
is $\sim$10$^{44}$~erg/s, blazars are spectacular objects, which 
are solely in \grays\ as luminous as thousands of galaxies, 
and can even switch on and off such luminosities on time scales 
of a day and less. 

Comparison of \gray\ emission with that in other observing bands 
seems to be most illuminating in the form of spectral energy 
distribution (SED) plots of $\nu$F$_{\nu}$ as a function of 
frequency. Such figures show the relative amounts of power
radiated in equal logarithmic frequency bands, 
and therefore provide the possibility to identify the major 
emission components. Two such spectra (for the blazars 3C~273 and 3C~279) 
are shown in Fig.~\ref{fig:blazars_spec}. 
3C~273 is bright at all wavelength regions and therefore 
is probably the best-studied blazar.  
First of all its SED proves that the high-energy emission
(X to \grays) is a significant part of the bolometric luminosity.
The spectrum shows
(probably) four maxima indicating different emission components
and processes, 
which are interpreted as synchrotron emission from relativistic
electrons in the radio- and far-IR band,
thermal emission from a dust torus in the IR and from an accretion
disk in the UV (\lq blue bump\rq), and inverse-Compton radiation
generated by relativistic electrons and soft photons at
X and \grays. 
In contrast to 3C~273 (z = 0.158), the simultaneous multifrequency spectra
of the blazar 3C~279 (z = 0.538) shows probably only two maxima. 
One is in the radio/infrared region and one is in the \gray\ band,
which, at least during \gray\ flaring events, 
clearly dominates the overall power output (Fig.~\ref{fig:blazars_spec}).
The spectrum is interpreted to be completely of non-thermal origin
with only two visible emission mechanisms at work: 
synchrotron emission from radio to UV/soft X-rays and
inverse-Compton emission from X-rays to \grays.
The thermal signatures are not visible. Probably they 
are outshone by the dominance of the non-thermal radiation. 

\subsection{TeV Blazars}

\begin{figure}[th]
\centerline{\epsfig{file=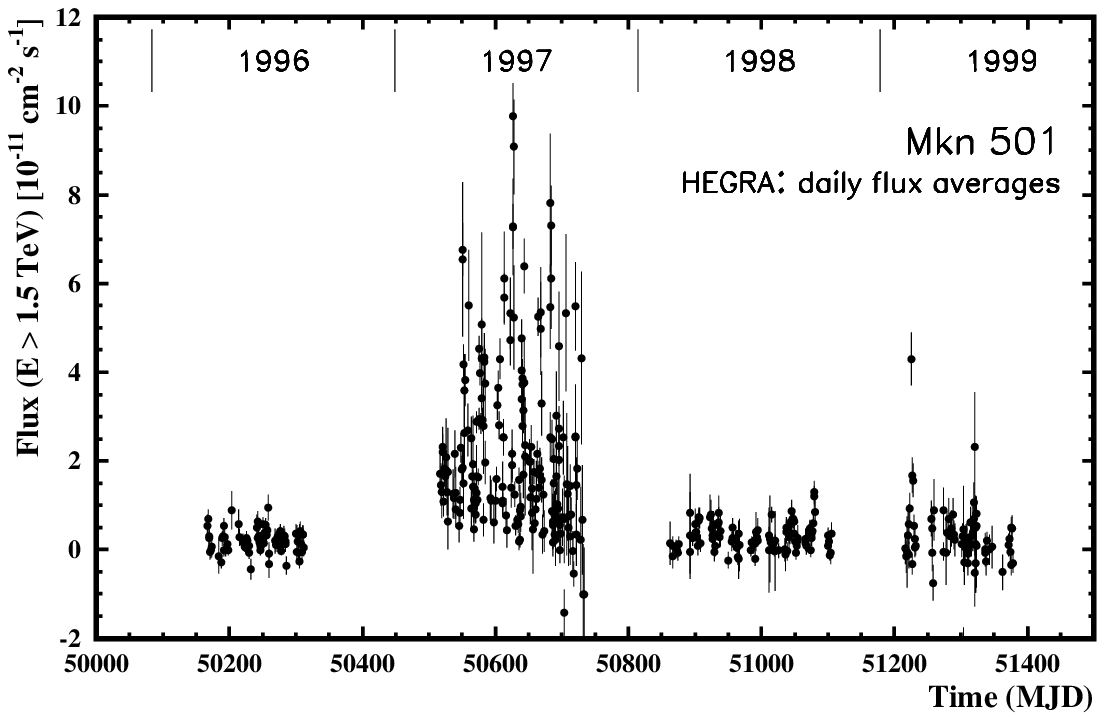,width=8.0cm,height=5.0cm,clip= }}
\centerline{\epsfig{file=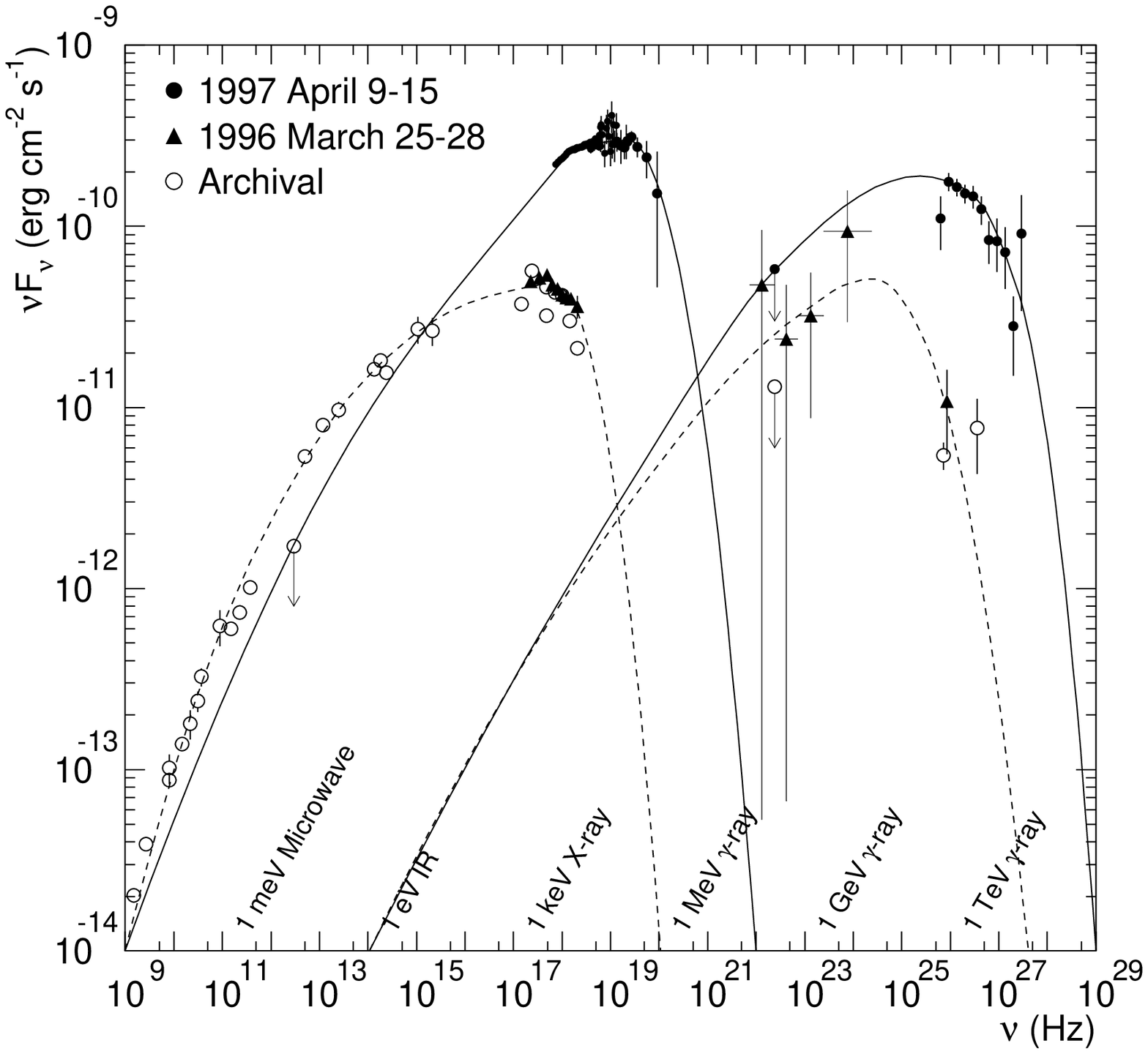,width=8.0cm,height=5.0cm,clip= }}
\caption
{The upper panel shows the long-term light curve of the TeV-emission 
($>$1.5~TeV) of the blazar Mkn~501. 
The daily flux averages as measured by the HEGRA telescope
during the years 1996 to 1999 are shown. 
Strong flaring activity is visible during the observational period in 1997.
(Credit for data: HEGRA Collaboration.)
\newline
The lower panel shows multifrequency spectra of Mkn~501
for two observational periods (TeV-quiet and flaring),
which are supplemented by archival data \protect{\citep{Catanese00}}.
The curves in the figures are spline functions and do not represent 
proper model fits. They are meant to guide the eye for the anticipated 
broad-band spectrum of Mkn~501. It is obvious that during the TeV flaring 
period in April 1997 the TeV flux as well as the X-ray flux 
increased significantly compared to the TeV quiescent period in March 1996.  
\label{fig:mkn501}}
\end{figure}

During the last decade a significant improvement in detector 
sensitivity was achieved for the Atmospheric Cherenkov Telescopes 
(ACTs) which measure photons at energies between $\sim$300~GeV and 10~TeV,
in the so-called VHE (very high energy) band.
Up to now, 6 sources \citep[e.g.,][]{Weekes00} 
have been reported to emit at these energies. The most prominent ones 
are Markarian (Mkn)~421 and Mkn~501 which were both detected 
several times by several 
groups (e.g., Whipple collaboration, HEGRA collaboration). 
Due to these multiple and significant detections both sources
are well studied 
at TeV energies: Both blazars are time variable and show strong 
flaring activity with flux-doubling times 
as low as 15 minutes \citep{Gaidos96}. 
Fig.~\ref{fig:mkn501} shows the long-term light curve of Mkn~501, which 
indicates quiescent and flaring periods. 
The VHE spectra are different for both sources. The one of Mkn~421 shows 
no obvious cutoff either at the lower or the upper end
\citep[e.g.,][]{Krennrich99, Aharonian99}, and therefore
is well represented by a power-law model. The spectrum of Mkn~501
in contrast, clearly shows a curved shape \citep[e.g.,][]{Samuelson98} 
during strong flaring 
events (see Fig.~\ref{fig:mkn501}). This spectral difference is surprising
because both sources belong to the same class of objects, 
X-ray selected BL~Lacs (XBLs), and have roughly the same distance. 
Most likely the different spectra point to different intrinsic 
properties. 

Multifrequency observations of both Mkn~421 and Mkn~501
indicated a connection between the TeV and X-ray band
\citep[e.g.,][]{Buckley96, Catanese97}.
Flares seem to occur simultaneously, but with larger amplitudes at
TeV-energies. Surprisingly, both sources are weak \gray\ emitters
in the MeV/GeV range. This is well illustrated by broad-band spectra,
which during TeV-flares (making the measurement of a spectrum possible 
at these energies)
show the same two-peak spectra as the usual EGRET-type blazars.
However, both maxima are shifted to higher energies (Fig.~\ref{fig:mkn501}).
One appears in the X-ray or occasionally even in the hard X-ray band 
and the other one in the TeV-range.

\begin{figure}[th]
\centering
\epsfig{figure=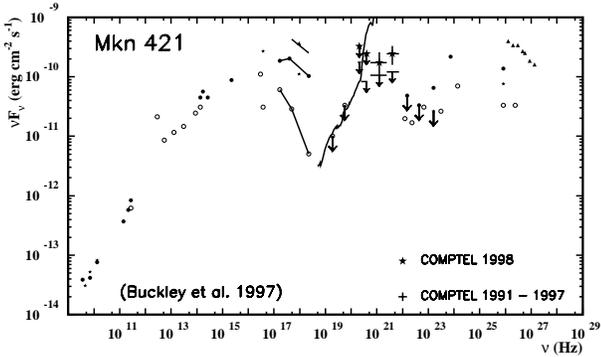,width=8cm,clip=}
\vspace{-0.3cm}
\caption{
Multifrequency data of Mkn~421 for different epochs. 
The variability at X-rays and TeV \grays\ is visible. 
The 3$\sigma$ sensitivity lines for an observation time 
of 10$^{6}$ seconds of the INTEGRAL SPI instrument is
included. The data, apart from the COMPTEL ones, are from 
\protect{\citet{Buckley96}}. The COMPTEL data are from 
\protect{\citet{Collmar99b}}.  
\label{fig:mkn421}}
\end{figure}

INTEGRAL measurements will cover the decreasing part of the synchrotron 
spectrum and the \lq flux valley\rq\ between the two peaks
(Fig.~\ref{fig:mkn421}).
Correlated measurements with the new-generation ACTs, which will have 
lowered energy thresholds of $\sim$50~GeV and which will become 
operational in 2001/2, have the potential to improve our knowledge 
on the emission processes in these sources. 
In particular, they will distinguish between intrinsic source properties 
and photon propagation effects, i.e., whether the curved
TeV spectra of Mkn~501 are intrinsically generated or are due to
energy-dependent photon absorption between the sources and us.

\subsection{Interpretation}

The large \gray\ luminosities of blazars together with the
 observed short-term variability imply a
highly compact emission region of the \gray\ radiation. 
For an isotropic luminosity of 10$^{48}$~erg/s 
and a radius of 1 light day,
as observed in 3C 279 for example, an optical depth for pair production
of $>$200 is derived. This simply means that \grays\ cannot escape without
generating electron-positron pairs. Nevertheless, these \grays\
 are observed!  This contradiction is resolved if emission
from a relativistic jet is considered instead of central isotropic
emission which was assumed in the example above.
A jet origin of the observed \grays\ is consistent with 1)
the fact that for many blazars superluminal
motion has been observed, which is indicative of jets with
a small offset angle to our line-of-sight,
and 2) the redshift distribution of these sources, which shows that
the source distance is not the critical parameter for their
detection, because \gray\ blazars are observed far out into the
universe up to redshifts of z=2.3. A relativistic jet origin of
the \grays\ would imply beamed emission.
The beamed emission has another effect: the observed
\gray\ luminosity overestimates the internal generated luminosity
by a factor of up to ~10$^{4}$ due to relativistic Doppler boosting
of the emitted photons.
Therefore internally the optical depth for pair production is
well below 1, which is in accord with the observation of 
these \grays. 

Since the CGRO discovery of \gray\ blazars,
the origin of their \gray\ emission has been 
widely discussed. 
As concluded above, isotropic core emission has to be excluded. 
The favoured scenario to explain the blazar spectral
continua is that we  are viewing nearly
along the axis of a relativistically outflowing plasma jet which has
 been ejected from an accreting supermassive black hole.
This broadband radiation is thought to be produced by non-thermal
electron synchrotron radiation in outflowing plasma blobs 
or shocked jet regions.
The high-energy blazar
continuum emission appears to constitute a distinct second component
in the broadband spectral energy distribution of blazars.
Two classes of models have been proposed to explain the blazar
\gray\ radiation, in which either leptons or hadrons are the
primary accelerated particles, which then radiate
directly or through the production of secondary particles
which in turn emit photons.

\begin{figure}[t*]
\centering
\epsfig{figure=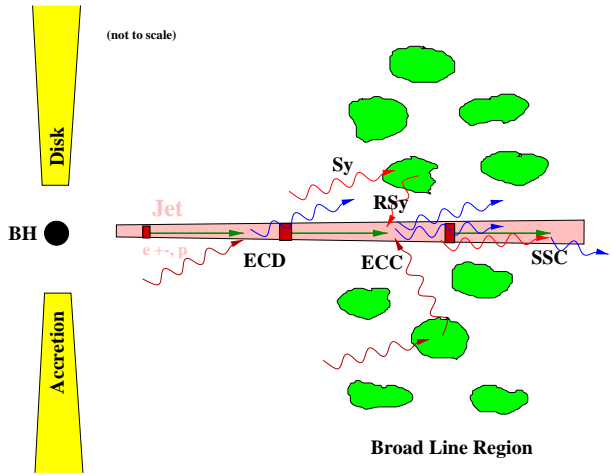,width=8cm,clip=}
\caption{
Sketch (not to scale) of a model for the central part of a 
radio-loud AGN. Possible emission processes for the leptonic 
models are indicated. The \grays\ are generated by 
inverse-Compton upscattering of soft photons off 
relativistiv electrons (inside the jet blobs).
The different acronyms
refer to different populations of soft photons
(ECD: upscattering of photons coming directly from the accretion disk;
ECC: upscattering of accretion disk photons isotropized by the
 broad-line region clouds;
SSS: upscattering of self-generated (inside jet) synchrotron 
photons;
RSy: upscattering of synchrotron photons reflected by the
 broad-line region clouds \protect{\citep{Ghisellini96}}.    
\label{fig:sketch}}
\end{figure}

In leptonic models (Fig.~\ref{fig:sketch} sketches the central part
of an AGN with a leptonic emission scenario),
 the \gray\ emission of blazars
is produced by non-thermal relativistic leptons 
(electrons and positrons) which scatter soft photons to
\gray\ energies via the inverse-Compton
(IC) process. These leptonic models come in two flavors 
depending on the nature of the soft photons.
The synchrotron-self Compton models (SSC) assume the SSC process
 to be dominant \citep[e.g.,][]{Maraschi92, Bloom96}.
 In this process
the relativistic electrons moving along the magnetized jet
generate synchrotron photons which are
boosted by the same relativistic electron population to
\gray\ energies via the IC-process. The IC-spectrum
follows to first order the shape of the synchrotron spectrum
(just shifted to higher energies), which explains the observed
spectral turnover at MeV-energies in the \gray\ spectra.

The so-called external Compton scattering (ECS) models
consider a different origin for the soft photons.
They assume the soft X-ray and UV-photons which  are radiated
from the accretion disk directly into the jet
\citep[e.g.,][]{Dermer93} or scattered into
the jet by the broad line region clouds 
\citep[e.g.,][]{Sikora94} to be the soft target
photons for the relativistic jet electrons. The ECS-models
also can reproduce the broad-band non-thermal spectral shape
of the \gray\ blazars. These models explain the spectral
bending at MeV-energies by the so called
incomplete Compton cooling of the electrons. 
When a blob of relativistic electrons is injected into the
jet, a power-law shaped IC-spectrum is generated
with a low and high-energy cutoff in photon energy
which correspond to
the low - and high-energy cutoffs in the electron spectrum.
Because the high-energy photons cool first,
the high-energy cutoff in the IC-spectrum moves towards
lower energies with time. The electron cooling by the
IC-process stops when the blob has moved out into regions
where the photon field becomes too thin to maintain this process.
Integrating the individual spectra over time 
generates the spectral turnover at MeV energies observed by CGRO in time-averaged (days to weeks) \gray\ spectra.      

Models have also been proposed in which accelerated
hadronic particles (mainly protons) carry the bulk 
of the energy \citep[e.g.,][]{Mannheim92, Bednarek93}.
Because protons do not suffer severe radiation losses,
they can be accelerated up to energies of 10$^{20}$eV, reaching
the thresholds for  photo-pion production.
In these processes the protons transfer energy into
photons, pairs, and neutrinos via pion production.
The photons and pairs are reprocessed and initiate a
cascade through inverse-Compton and synchrotron processes
to form a power-law photon spectrum in the end. Electrons and positrons generated via charged pion decay will be accompanied by
energetic neutrino production. The detection of a strong neutrino
flux from blazar jets would definitively
identify hadrons as the primary accelerated particles.

\section{Seyferts}
\subsection{Radio-quiet Seyferts}

\begin{figure}[th]
\centering
\epsfig{figure=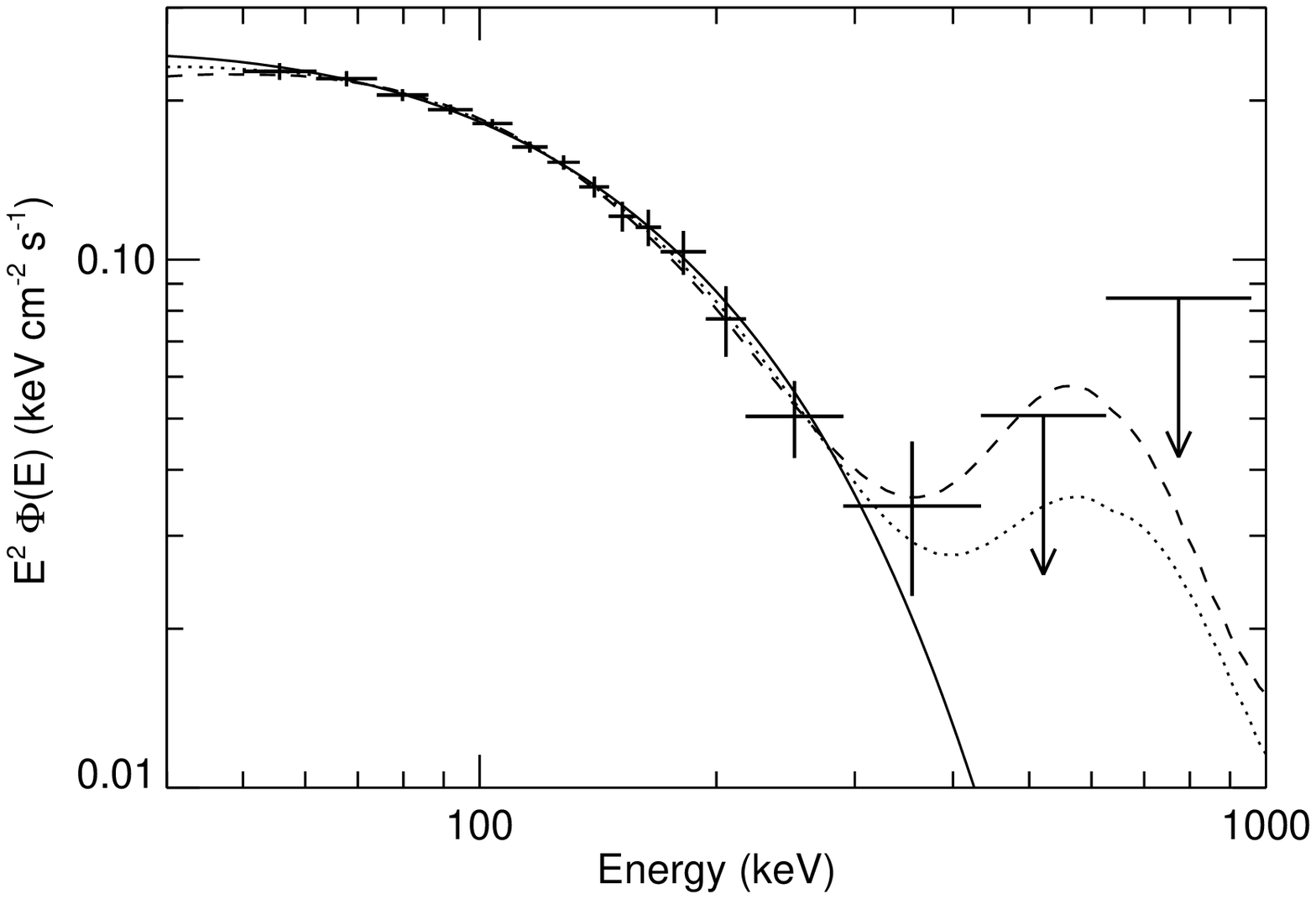,width=8cm,clip=}
\epsfig{figure=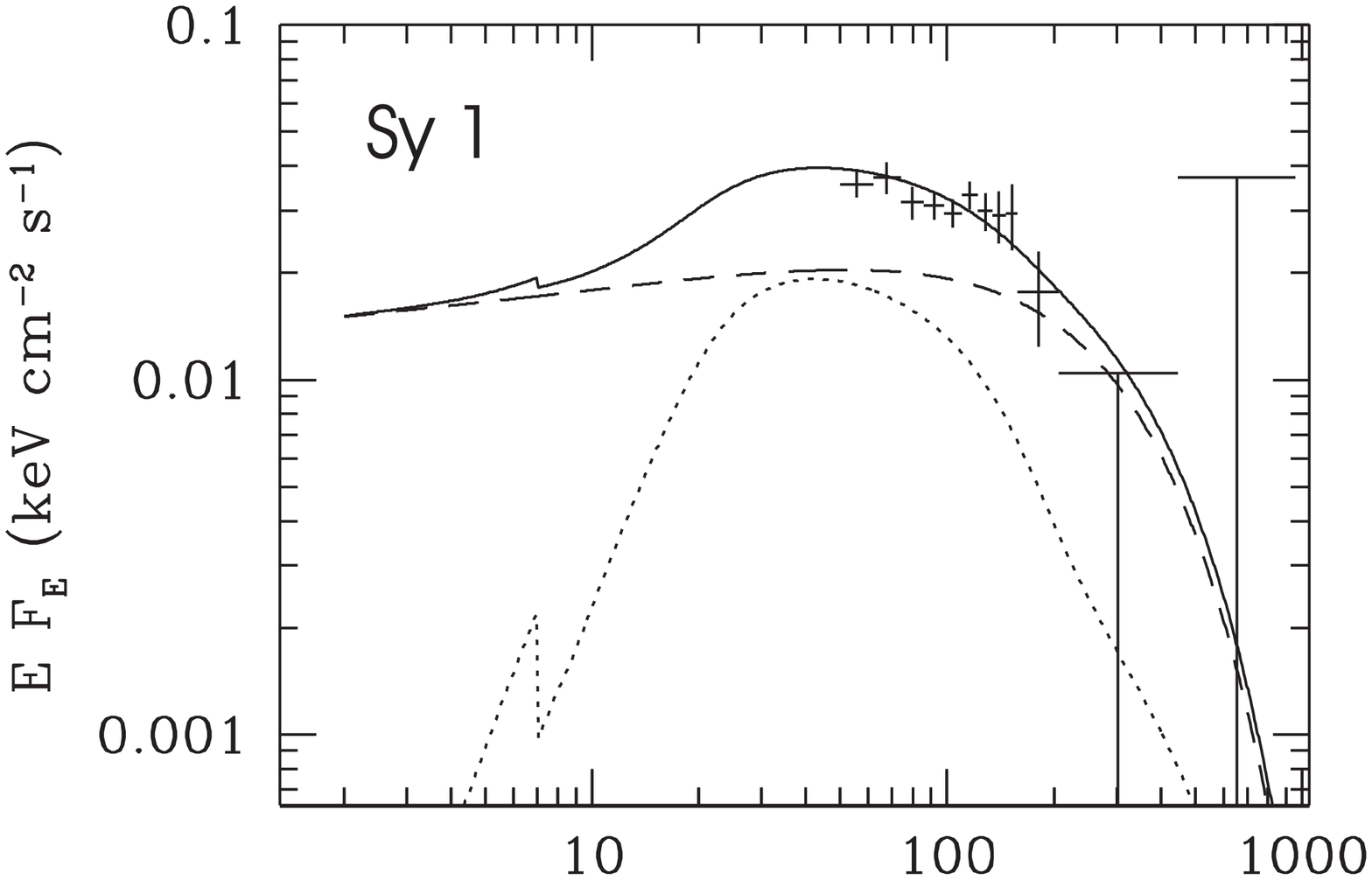,width=8cm,clip=}
\vspace*{-0.5cm}
\caption{
Upper panel: 
The average OSSE spectrum (50~keV to $\sim$1~MeV) of NGC~4151, 
the brightest Seyfert galaxy, is shown. The solid curve represents 
the best-fit thermal Comptonization plus Compton-reflection model.
The dashed curve corresponds to the best-fit hybrid,
thermal/non-thermal model, and the dotted curve shows a model
with a 15\% non-thermal fraction, i.e. 15\% of the total power is 
going into electron acceleration \protect{\citep[from][]{Johnson97}}.   
\newline
Lower panel: Average spectrum of the Seyfert 1 class as measured with OSSE.
The best-fit model (solid line) together with the individual 
model components are shown throughout the X-ray and hard X-ray band.
The dashed curve represents the best-fit 
thermal Comptonization spectrum and the dotted curve the 
reflected component including a flourescent Fe K$\alpha$ 
line \protect{\citep[from][]{Johnson97}}. 
\label{fig:OSSE_Sys}}
\end{figure}

Radio-quiet Seyferts, the standard Seyfert classes 1 and 2, 
are nearby 
AGN without strong jets. Prior to CGRO Seyferts were considered
to be promising candidates for \gray\ emission. 
An extrapolation of their hard power-law spectra measured at X-rays
(2 - 20~keV) and hard X-rays ($>$20~keV) into the \gray\ domain
($\geq$500~keV) resulted in large
\gray\ fluxes, and subsequently in strong and significant 
sources for the CGRO instruments. 
Now, after the end of the CGRO mission, roughly 25 Seyferts 
are detected by the OSSE \citep[e.g.,][]{Johnson97}
and BATSE experiments \citep[e.g.,][]{Bassani96, Maliza97}. 
The detection occur mainly in the hard X-ray band between 
$\sim$50 and $\sim$300~keV. No Seyfert has been \lq seen\rq\ 
at 1~MeV and above, making them rather hard X-ray than 
\gray\ sources \citep[e.g.,][]{Maisack95}. 

The brightest and  best-studied object is NGC~4151, from which
much of the current knowledge is derived. OSSE measured 
time variability of up to a factor of $\sim$2 on long time scales
(years) and up to 25\% on a day-to-day basis \citep{Johnson97}. 
Its spectrum at these energies is well described by a power law 
which cuts off exponentially (Fig.~\ref{fig:OSSE_Sys}): 

\begin{equation}
dN/dE \propto E^{-\alpha} \times e^{-\frac{E}{E_{C}}}  \enspace ,  \label{none}
\end{equation} 

where $\alpha$ is the photon index, and $E_{C}$ is the cutoff energy. 
Such a mathematical shape approximates a thermal Comptonization spectrum. Best-fit values of $\alpha$ of $\sim$1.6 and $E_{C}$ of
$\sim$100~keV are derived. 
In particular, no (strong) 511~keV pair--annihilation emission is found,
which severely restricts emission models for Seyferts. 
In summary, the OSSE data on NGC~4151 suggest that its high-energy 
emission is well described by thermal processes.
An upper limit of 15\% for the contribution of non-thermal
processes is derived \citep{Johnson97}.  
The results on NGC~4151 are consistent with the results 
of analysing the sum of several weaker sources (Fig.~\ref{fig:OSSE_Sys}). 

To test emission models for the high-energy emission of Seyferts, 
combined X-ray and hard X-ray spectra have been compiled.
Such spectra from radio-quiet Seyferts are well described by a
power--law shape at X-rays which cuts off around 100~keV
(as described above), and which 
is interpreted to be the Comptonized emission from a hot plasma. 
This primary component is modified by the presence of nearby matter, i.e.
a reflection component becomes visible at energies above $\sim$10~keV
(Fig.~\ref{fig:OSSE_Sys}). 
A correlation of the strength of the Compton reflection component
with the spectral index of the primary emission, the Comptonized one,  
is found by \citet{Zdziarski00}. This implies that the Comptonizing 
plasma and the cold medium are connected. \citet{Zdziarski00} 
suggest that the cold medium provides the soft photons which are 
Compton upscattered in the hot plasma.  
 
\begin{figure}[th]
\centering
\epsfig{figure=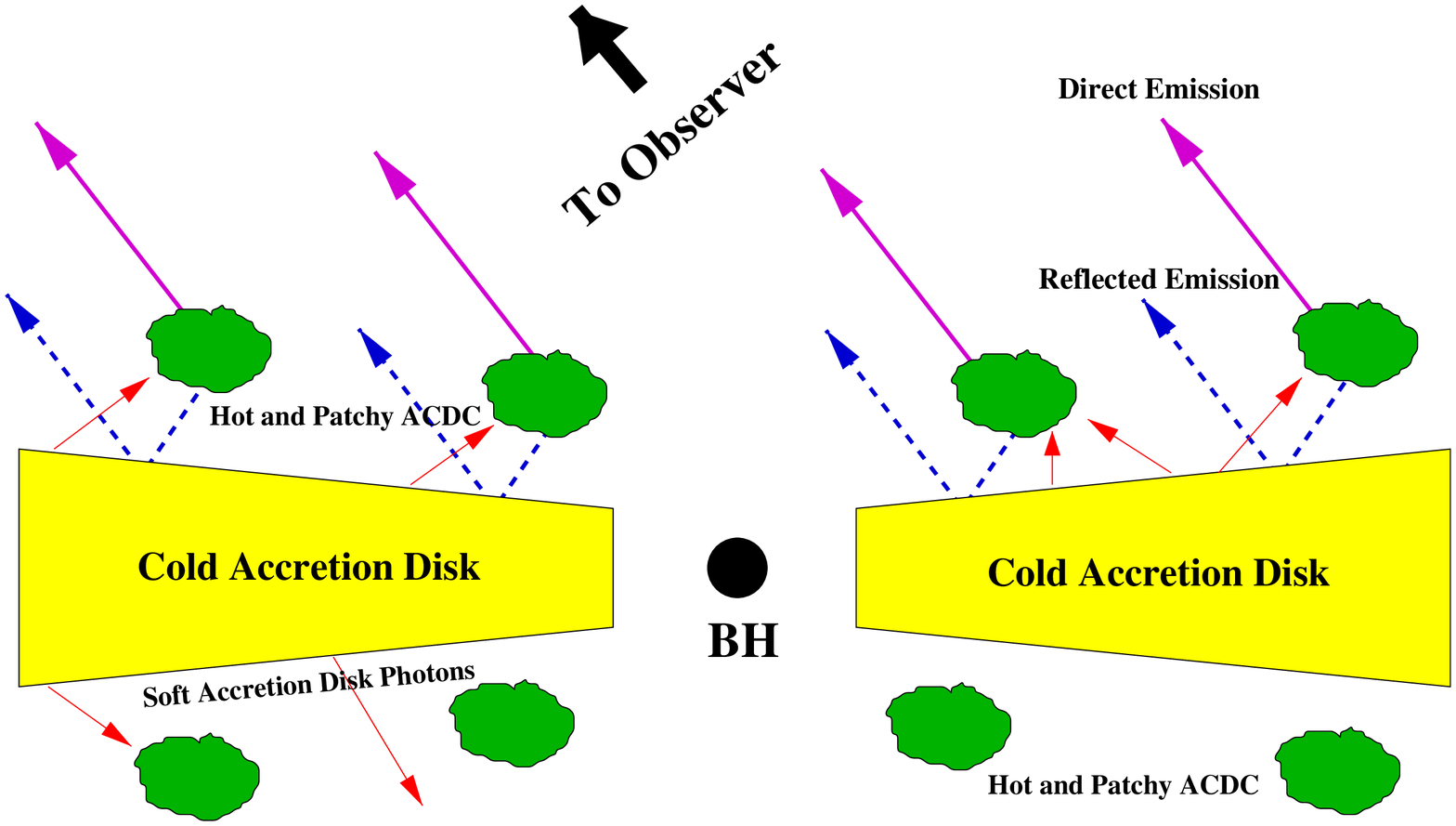,width=8cm,clip=}
\epsfig{figure=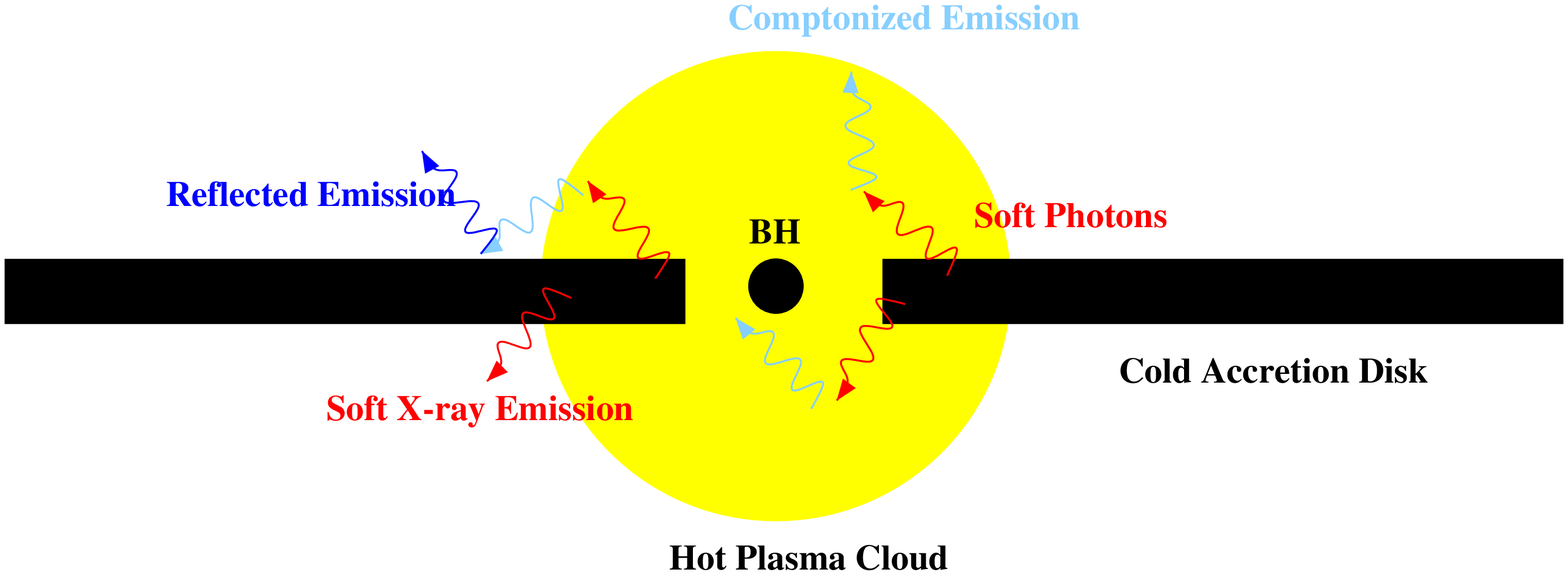,width=8cm,clip=}
\caption{
Sketches of possible geometries of the central and emitting 
region of radio-quiet Seyfert galaxies. The two emission
components, direct (Comptonized) and reflected emission, 
are indicated. Two scenarios \protect{\citep{Zdziarski00}}
are consistent with current data: an active, non-homogeneous,
corona above a cold accretion disk (upper panel) and
central core emission (lower panel) from a hot plasma cloud
(e.g., a ADAF-type inner disk) overlapping with a cold medium 
(e.g., a cold outer disk). 
\label{fig:sketch_Sys}}
\end{figure}

According to current data, two emission geometries 
appear to be possible (Fig.~\ref{fig:sketch_Sys}): 
a patchy corona above a cold accretion disk, 
and a hot accretion disk with an overlapping cold medium, 
e.g., a hot inner ADAF-type (advection dominated accretion flow)
disk and a cold outer one. 
Both geometries fulfill the requirement that the reflection
component subtends a solid angle less than 2$\pi$.

INTEGRAL will provide improved measurements of Seyferts in the hard 
X-ray to \gray\ band. This will -- in particular --
result in improved constrains on the contribution of non-thermal
processes (i.e., the level of particle acceleration 
taking place in these sources), and in more sensitive searches
for pair annihilation.    

\subsection{Radio-loud Seyferts}

\begin{figure}[th]
\centering
\epsfig{figure=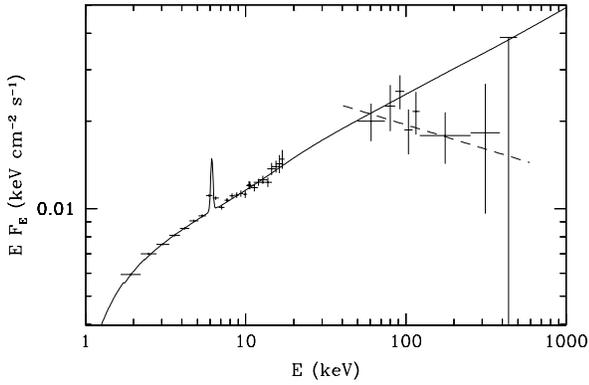,width=8cm,clip=}
\caption{
The average X-ray to \gray\ spectrum of radio-loud Seyferts
from GINGA ($\sim$2 - 20~keV) and OSSE ($>$50~keV)
 observations of 3 sources each. 
The GINGA X-ray spectrum is fitted with a power-law incident 
continuum plus Compton reflection (solid line), and the 
OSSE hard X-ray spectrum with a single power-law model 
\protect{\citep{Wozniak98}}. 
\label{fig:loud_Sys}}
\end{figure}

The class of radio-loud Seyferts, which are also called broad-line 
radio galaxies, is populated by only a few sources.
These sources show jets and one member -- 3C~120 -- even superluminal 
motion. 
The spectra at X-ray energies are of power-law shape with an average 
photon index ($\alpha_{ph}$) of $\sim$1.7 \citep{Wozniak98}.
Like the radio-quiet Seyferts these spectra break around 50 - 100~keV 
to a softer shape, which however, unlike the radio-quiet Seyferts, does 
not show an obvious cutoff (Fig.~\ref{fig:loud_Sys}).
It is consistent with a power-law shape 
of $\alpha_{ph}$ of $\sim$2.2 \citep{Wozniak98}, which might indicate 
the presence of non-thermal Compton scattering. 
However, the statistical significance of the OSSE spectra of these 
sources is well below those of the radio-quiet Seyferts, and
therefore this conclusion is only tentative. 

More sensitive hard X-ray measurements of these sources 
are needed to draw improved conclusions. INTEGRAL can provide
such measurements, and therefore might be able to decide 
whether radio-loud Seyferts have a different or the same type
of emission scenario than radio-quiet ones. 
If non-thermal emission would be important in these sources, 
then they would be a link between the radio-quiet Seyferts and 
the blazars.

\section{Radio Galaxies}

\begin{figure}[th]
\centering
\epsfig{figure=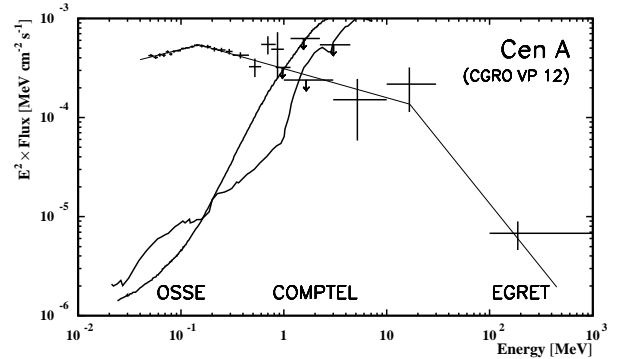,width=8cm,clip=}
\caption{
Broad-band spectrum of Centaurus~A from X to GeV \gray\ energies.
The spectral break at hard X-rays ($\sim$150~keV) is evident. 
To match the EGRET flux a second break is necessary 
\protect{\citep{Steinle98}}.
The 3$\sigma$ sensitivity lines for an observation time 
of 10$^{6}$ seconds of the INTEGRAL instruments IBIS and SPI
are included.  
\label{fig:cen-a}}
\end{figure}

Centaurus~A (Cen~A), the closest ($\sim$3.5~Mpc) active galaxy, 
is detected from radio to GeV \grays. 
It is a unique and interesting object, because it is the 
only non-blazar AGN detected at MeV-energies and above.
Its X to \gray\ spectrum can be described by different power-law
shapes, showing a spectral steepening towards higher energies
(Fig.~\ref{fig:cen-a}). In hard X-rays, 
the spectrum has the canonical hard power--law shape with
photon index $\sim$1.7, 
like the radio-loud and radio-quiet Seyferts. 
It breaks at $\sim$150~keV to a softer power--law shape
($\alpha_{ph}\sim$2.3) which holds up to $\sim$30~MeV, and
has to break again ($\alpha_{ph}\sim$3.3)
to match the weak EGRET flux above $\sim$100~MeV \citep{Steinle98}.
Since there is no exponential cutoff visible in hard X-rays,
resembling the spectra of the radio-loud Seyferts,
non-thermal processes are probably operating in this object
at these energies. 
The emission around 1~MeV and above is generally considered
to be of non-thermal origin.

The emission scenario of Cen~A is unclear.
It is still an open question whether we \lq see\rq\ thermal and/or 
non-thermal emission, and therefore it is still open
whether the origin of the emission is located in the 
nucleus and/or the jet. This might be changing with respect 
to energy. 
Because Cen~A is viewed from the \lq side\rq
 -- the angle between its jet axis and our line--of--sight
is $\sim$70\deg\ \citep[e.g.,][]{Jones96} -- the hypothesis of a 
\lq misdirected\rq\  blazar can be considered.
In this case we might observe the intrinsically ({\bf not}
 Doppler-boosted) 
generated \gray\ luminosity, which is of the order of
10$^{42}$~erg/s at MeV energies. 
This MeV luminosity -- if viewed close to \lq jet on-axis\rq\ 
and subsequently Doppler
boosted by a factor of $\sim$10$^{3}$ - 10$^{4}$ -- would place Cen~A
in luminosity near the weaker blazars detected by COMPTEL. 

\section{Summary}

Substantial progress has been made in the study of AGN at \gray\ energies 
during the last decade. Before the launch of CGRO only 4 sources 
had been reported to emit detectable \grays\ around $\sim$500~keV 
and above, which led to a vague and patchy knowledge about their 
emission properties and scenarios. Now, after the end of
the CGRO mission, more than 100 sources are detected in the energy 
band from hard X-rays ($>$50~keV) to  
GeV \grays\ by the different CGRO experiments.
Even at VHE \grays\ at least two AGN 
are convincingly detected and well studied now. 
Due to these many observations and source detections, a consistent
picture has emerged during these years, which is generally understood. 
However, most details are still unknown.  

\begin{figure}[th]
\centering
\epsfig{figure=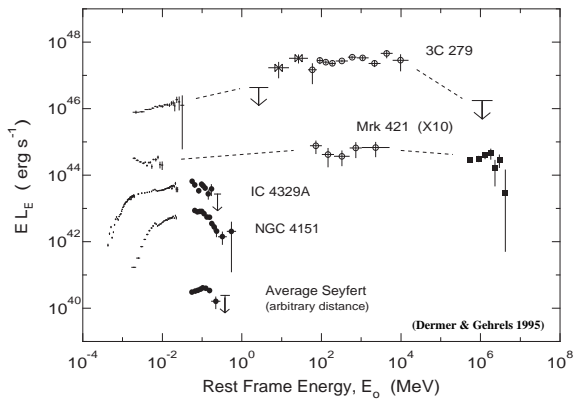,width=8cm,clip=}
\caption{
Multiwavelength power spectra for the \lq average\rq\ Seyfert galaxy,
the bright Seyferts NGC 4151 and IC 4329A, and
the blazars Mkn~421 and 3C~279.
 The figure clearly shows that in contrast to both blazars,
 the high-energy Seyfert spectra cut off around 100~keV,
 indicating thermal instead of non-thermal emission processes
\protect{\citep[from][]{Dermer95}}.
\label{fig:de_ge}}
\end{figure}

The general picture is, that radio-loud AGN, especially blazars,
can occasionally be strong emitters of \gray\ radiation, 
which is detectable from MeV to GeV, and for a few sources even to 
TeV energies. It is generally agreed that this radiation emerges 
from a relativistic plasma jet and is generated via
inverse Comptonization of soft photons by relativistic electrons.
On the other hand,
radio-quiet sources show the signs of thermally Comptonized spectra, 
which cut off at energies around $\sim$100~keV. This emission is 
thought to be generated in the central AGN region, i.e. in the 
vicinity of the putative supermassive black hole.   
This general picture is shown in Fig.~\ref{fig:de_ge}, which compares 
the high-energy luminosity spectra (assuming isotropic emission) 
of different sources belonging to different source classes. 
This figure, in fact, allows one to compare the radiated power
as a function of energy. 
The different spectral behaviour of blazars,
having non-thermal jet emission, and Seyferts,
showing thermal emission from the accretion disk, is obvious. 
Additionally, the figure shows that the BL Lac type blazar Mkn~421
even maintains its emission into the TeV range.

The recent progress in this field has not only provided new results
and answers, but as usual has also posed many new questions. 
 \lq What causes the \gray\ flares in blazars?\rq,  
\lq What kind of emission -- thermal and/or non-thermal --
do we observe in Cen~A and radio-loud Seyferts?\rq,
are only two examples. They can and will be \lq attacked\rq\ by future 
missions which are already on the horizon, like INTEGRAL, AGILE 
and GLAST, to improve on the current general picture.

\end{document}